\documentstyle[pra,preprint,aps]{revtex}
\tightenlines
\begin{document}
\title{Precise fine-structure and hyperfine-structure 
measurements in Rb}
\author{Ayan Banerjee, Dipankar Das, and Vasant 
Natarajan\thanks{Electronic address: 
vasant@physics.iisc.ernet.in}}
\address{Department of Physics, Indian Institute of Science, 
Bangalore 560 012, INDIA}

\maketitle

\begin{abstract}
We demonstrate a new technique for measuring the absolute 
frequencies of atomic transitions. The technique uses a 
ring-cavity resonator whose length is calibrated using a 
reference laser locked to the $D_2$ line in $^{87}$Rb. The 
frequency of this line is known to be 384 230 484.468(10) 
MHz. Using a second laser locked to the $D_1$ line of Rb, we 
measure the frequencies of various hyperfine transitions in 
the $D_1$ and $D_2$ lines with a precision of 30 kHz. We 
obtain the following values: 120.687(17) MHz and 406.520(25) 
MHz for the $5P_{1/2}$ hyperfine constant $A$ in $^{85}$Rb 
and $^{87}$Rb; 377 107 385.623(50) MHz and 377 107 
463.209(50) MHz for the $D_1$ line in $^{85}$Rb and 
$^{87}$Rb; and 384 230 406.528(50) MHz for the $D_2$ line in 
$^{85}$Rb. This yields the fine-structure interval and the 
isotope shifts. The precision obtained is a significant 
improvement over previous measurements.

\end{abstract}

Precise measurements of atomic energy levels continue to 
play an important role in the development of physics. The 
energy levels of alkali atoms are particularly important 
because of their widespread use in laser-cooling 
experiments, ultra-cold collision studies, photoassociation 
spectroscopy, atomic tests of parity violation, and more 
recently in Bose-Einstein condensation. For example, 
measurement of the $D_1$ line in Cs can lead to an improved 
value for the fine-structure constant $\alpha$ \cite{URH99}. 
In addition, knowledge of the fine-structure interval in the 
$D$ line of alkali atoms is important for several reasons, 
{\it e.g.\ }in the study of atomic collisions, astrophysical 
processes, and relativistic calculations of atomic energy 
levels.

In this Letter, we demonstrate a new technique for precisely 
measuring the energy levels of atoms. The technique combines 
the advantages of using diode lasers to access atomic 
transitions with the fact that the absolute frequency of the 
$D_2$ line in $^{87}$Rb has been measured with an accuracy 
of a few kHz \cite{YSJ96}. A stabilized diode laser locked 
to the $D_2$ line of $^{87}$Rb is used as a frequency 
reference along with a ring-cavity resonator whose length is 
measured with the reference laser. For a given cavity 
length, an unknown laser locked to an atomic transition has 
a small frequency offset from the nearest cavity resonance. 
This offset acts as a vernier scale and is combined with the 
cavity mode number to obtain a precise value for the 
frequency of the unknown laser. We have used this method to 
make precise measurements of energy levels in the $D$ lines 
of Rb. This yields the hyperfine structure in the $5P_{1/2}$ 
state and the $D_1-D_2$ fine-structure interval. We 
demonstrate a precision of 30 kHz, which is two to three 
orders of magnitude better than the typical accuracy of 
energy-level tables \cite{MOO71}. While we have measured 
energy levels only in Rb, the technique is more general and 
should prove useful in other alkali atoms (such as Li, K, 
and Na), alkali-like ions, and indeed any system where the 
transitions are accessible with tunable lasers.

Similar measurements have been done previously using a 
stabilized HeNe laser as the frequency reference and a linear 
Fabry-Perot resonator \cite{BGR91}. Our use of a Rb-stabilized 
diode laser as the reference has the primary advantage that 
there are several hyperfine transitions that can be used for 
locking the laser. This enables us to check different sources 
of systematic errors. Our use of a ring-cavity resonator also 
has several advantages. The main problem with a linear 
resonator is that it can cause unwanted feedback into the 
diode laser and destabilize it. By contrast, the ring-cavity 
resonator has a traveling wave inside the cavity and there is 
no possibility of feedback into the laser. The other 
advantages of the ring cavity are: i) The cavity provides 
three output beams that can be used to measure the cavity 
signal. ii) The mode structure inside the cavity is elliptical 
which makes it easier to mode-match the elliptical output of 
the diode laser into the cavity. iii) For a given cavity 
length, the design is more compact and the cavity is easily 
temperature stabilized to increase its passive stability.

The schematic of the experiment is shown in Fig.\ 1. The 
experiment uses two diode lasers that are both locked to 
hyperfine transitions in Rb using Doppler-free 
saturated-absorption spectroscopy. Laser 1 is on the $D_1$ line (whose 
frequency has to be measured) and Laser 2 is on the $D_2$ 
line (which acts as the frequency reference). The output of 
the two lasers is fed into the ring-cavity resonator. In 
general, the length of the cavity is such that the cavity is 
not in resonance with either laser. The cavity length is 
adjusted using a piezo-mounted mirror to bring it into 
resonance with the wavelength of Laser 1. The cavity is then 
locked to this length in a feedback loop. However, Laser 2 
will still be offset from the cavity resonance. This offset 
is accounted for by shifting the frequency of the laser 
using an acousto-optic modulator (AOM) before it enters the 
cavity. The error signal between the shifted frequency of 
Laser 2 and the cavity resonance is fed back to the AOM 
driver which locks the frequency of the AOM at the correct 
offset. The frequency of the AOM is read using a frequency 
counter and thus the exact offset between the two lasers is 
determined. The absolute frequency of Laser 2 is known with 
10 kHz accuracy \cite{YSJ96}, therefore, once the cavity 
length (or mode number) is known, the frequency of Laser 1 
is determined very precisely.

The relevant low-lying energy levels for the two isotopes of 
Rb are shown in Fig.\ 2. The various hyperfine levels in the 
ground and excited states are given along with their energy 
shifts from the unperturbed state. Knowledge of the shift is 
important in obtaining the hyperfine-free transition 
frequency. The ground-state shifts are known with sub-kHz 
accuracy \cite{AIV77}. The shifts in the $5P_{3/2}$ state 
have been measured with an accuracy of 50 kHz for $^{85}$Rb 
\cite{RKN02} and with an accuracy of 10 kHz for $^{87}$Rb 
\cite{YSJ96}. However, for the $5P_{1/2}$ state, the values 
have large error bars of order 1 MHz \cite{AIV77}. In this 
work, we have improved these values by one to two orders of 
magnitude.

The diode lasers are standard external-cavity lasers 
stabilized using optical feedback from a piezo-mounted grating 
\cite{BRW01}. Large scans of the laser frequency are possible 
by changing the angle of the grating. The current through the 
diodes is dithered at a frequency of 10--50 kHz to produce the 
error signal needed for locking. The ring cavity consists of 
two plane mirrors and two concave mirrors in a bow-tie 
arrangement. One of the plane mirrors is partially reflecting 
(97\%) and is used to couple light into the cavity. The other 
plane mirror is mounted on a piezoelectric transducer and is 
used to adjust the cavity length electronically. The two 
concave mirrors have radius of curvature of 25 mm. They are 
placed 26.5 mm apart, while the optical path length between 
them through the plane mirrors is 200 mm. The angle of 
incidence on all mirrors is 15$^{\circ}$. The mirrors are 
mounted on a copper plate that is temperature controlled to 
$\pm 0.01^{\circ}$C using a thermoelectric cooler. The primary 
mode of the cavity has two beam waists, both of which are 
elliptical because of the non-zero angle of incidence on the 
curved mirrors. The large waist between the plane mirrors is 
used for efficient mode matching. The elliptic beams from the 
diode lasers are mode matched into the cavity using a convex 
lens.

The signal from a scan of Laser 2 is shown in Fig.\ 3. The 
upper trace is the power transmitted through the cavity, 
while the lower trace is the signal from the 
saturated-absorption spectrometer. The cavity length was fixed by 
locking it to Laser 1. It is seen from the figure that the 
hyperfine peaks are slightly shifted from the nearest cavity 
resonance. This is the shift that is accounted for by the 
AOM. The laser is locked to one of the peaks in the 
saturated-absorption spectrum using third-harmonic locking. 
This eliminates any systematic shift in the lock point due 
to the underlying Doppler profile in the spectrum. Note that 
the spectrum shown in the figure has the Doppler profile 
subtracted. Laser 1 is similarly locked to the 
saturated-absorption spectrum of the $D_1$ line using the 
third-derivative signal. However, locking to the cavity is to the 
standard first-derivative signal since this signal appears 
on a flat background.

The measurements rely on the fact that the cavity mode 
number for each laser is known exactly. For Laser 2, we 
determine it in the following manner. We first fix the 
cavity length by locking it to Laser 1. We then find the 
frequency offsets for two hyperfine transitions of Laser 2: 
$F=1 \rightarrow F'=(1,2)$ and $F=2 \rightarrow F'=(2,3)$. 
From Fig.\ 2, the frequency difference between these two 
transitions is exactly 6622.887(10) MHz. Since the cavity 
free-spectral range is about 1324 MHz, this corresponds to 
an increase in mode number by 5. Therefore, the two measured 
offsets are combined with the known shift to give a precise 
measurement of the cavity free spectral range. This 
determines the mode number uniquely. Once the mode number 
for Laser 2 is known, it is easy to find the mode number for 
Laser 1 as long as we know its frequency with an accuracy of 
about 50 MHz. This coarse measurement is done using a 
home-built wavemeter \cite{BRW01}. Note that the cavity length on 
resonance is an integer multiple of the laser {\it 
wavelength}, therefore, we have to account for the 
refractive index of air \cite{EDL66} when converting from 
frequency to wavelength.

The measurements of various hyperfine transitions are listed 
in Table I. Each value is an average over 50 individual 
measurements and the error given is the statistical error in 
the average. Systematic errors in the data can arise if 
there is any shift in the lock point of the lasers from the 
peak and optical-pumping effects from stray magnetic fields 
that change the lineshape of the spectrum. To check for such 
errors, we have repeated the same measurement using 
different hyperfine transitions for locking the reference 
laser. For example, measurements 1 and 2 in Table I are 
measurements of the same transition with the reference laser 
on the $F=1 \rightarrow F'=1$ and $F=2 \rightarrow F'=(1,3)$ 
transitions, respectively. If there is any systematic shift 
in the lock point of the reference laser, this would appear 
as a difference in the two values. Similarly, 3 and 4 are 
independent measurements of the same transition. Finally, 7 
and 8 are measurements of the $D_2$ line in $^{85}$Rb. For 
these two measurements, the role of the lasers is reversed 
since the unknown laser is on the $D_2$ line and Laser 1 
acts as the reference laser. We use our measured values for 
the $D_1$ line to obtain these values. These two 
measurements act as another check on the propagation of 
systematic errors since the difference between these values 
should be 31.705(50) MHz from Fig.\ 2. These checks on 
systematic errors enable us to estimate the size of the 
error in the individual frequencies to be no larger than 30 
kHz. In what follows, the errors quoted include 30 kHz of 
systematic error.

We first analyze our results in terms of the hyperfine 
structure of the $D_1$ line. Using the measured interval 
between the two hyperfine levels in the $5P_{1/2}$ state, we 
extract the value of the hyperfine constant $A$ in each 
isotope. The results are shown in Fig.\ 4. The values we 
obtain are 120.687(17) MHz for $^{85}$Rb and 406.520(25) MHz 
for $^{87}$Rb. Also shown in the figure is the comparison to 
earlier values: the recommended values of Arimondo et al.\ 
obtained from fitting to all the available data 
\cite{AIV77}, and the more recent results of Barwood et al.\ 
using a HeNe-stabilized Fabry-Perot interferometer 
\cite{BGR91}. Our results are consistent with the values of 
Arimondo et al.\ but have 2 to 3 orders of magnitude better 
accuracy. However, our values differ significantly from 
those reported by Barwood et al.\ even though their claimed 
error bars are slightly smaller. Note that their values also 
appear inconsistent with the recommended values.

We have also obtained the frequencies of the hyperfine-free 
$D$ lines and the fine-structure interval in Rb using the 
data from Table I. The results of this analysis are shown in 
Table II. The isotope shift is 77.586(70) MHz in the $D_1$ 
line and is 77.940(50) MHz in the $D_2$ line. The 
fine-structure interval is 7 123 020.905(70) MHz in $^{85}$Rb and 
7 123 021.259(50) MHz in $^{87}$Rb, yielding a value of 
0.354(90) MHz for the isotope shift in the interval. The 
accuracy of these values is significantly better than what 
has been reported earlier. The $D$-line frequencies in the 
published atomic energy tables have an accuracy of 300 MHz 
\cite{MOO71}. Barwood et al.\ \cite{BGR91} have reported a 
value of 377 106 271.6 MHz for the $D_1$ line of $^{85}$Rb 
($F=3 \rightarrow F'=3$) with an error of 0.4 MHz. Our 
values for the same transition (measurements 1 and 2 in 
Table I) are consistent but the accuracy is an order of 
magnitude better. For the fine-structure interval, we have 
earlier reported a value of 7 123 069(18) MHz using a scanning 
Michelson interferometer \cite{BRN01}, 
which overlaps with the current result at the $2.5\sigma$ 
level. To the best of our knowledge, there are no other 
measurements of the fine-structure interval. 

In conclusion, we have demonstrated a new technique for 
measuring the absolute frequencies of atomic transitions 
using a Rb-stabilized ring-cavity resonator. We have 
measured the frequencies of the $D$ lines in the two 
isotopes of Rb and significantly improved the precision of 
the fine-structure interval. We have also obtained improved 
values for the hyperfine structure in the $5P_{1/2}$ state. 
The main advantage of using a Rb-stabilized diode laser as 
the frequency reference is that it has several hyperfine 
transitions available for locking the laser, and the 
frequency shifts of these hyperfine levels are known very 
precisely. This is useful in measuring the mode number of 
the cavity. It also allows us to make several checks on 
systematic errors by making independent measurements of the 
same transition with different lock-points of the reference 
laser. We believe that this technique has the potential to 
improve the accuracy of atomic energy-level tables by 
several orders of magnitude. To demonstrate this, we plan to 
measure transitions in other atoms such as K and Yb.

We are grateful to Jhilam Sadhukhan for help with the 
measurements. This work was supported by research grants 
from the Board of Research in Nuclear Sciences (DAE), and 
the Department of Science and Technology, Government of 
India.

\begin{figure}
\caption{
Schematic of the experiment. The output of the two diode 
lasers is compared in the ring-cavity resonator. Laser 1 is 
the unknown laser locked to an atomic transition on the $D_1$ 
line. The cavity is locked to Laser 1. Laser 2 is the 
reference laser on the $D_2$ line. Its output is frequency 
shifted using the AOM before being fed into the cavity. The 
error signal between the shifted frequency and the cavity 
resonance is fed back to lock the AOM frequency. The frequency 
of the AOM gives the exact offset between the Laser 2 
frequency and the cavity resonance.
}
\label{schematic}
\end{figure}

\begin{figure}
\caption{
Rb energy levels. Shown are the relevant energy levels of 
$^{85}$Rb and $^{87}$Rb in the ground and excited states. The 
various hyperfine levels are labeled with the value of the 
total angular momentum quantum number $F$, and the number on 
each level is the hyperfine shift (in MHz) from the 
unperturbed state.
}
\label{rblevels}
\end{figure}

\begin{figure}
\caption{
Cavity modes. The upper trace shows the power in the cavity as 
the frequency of the Laser 2 is scanned and the lower trace 
shows the output of the saturated-absorption spectrometer. The 
length of the cavity is locked to Laser 1. The different 
hyperfine transitions in the $D_2$ line are at different 
offsets from the nearest cavity mode, which is accounted for 
using the frequency shift in the AOM. The two cavity modes are 
labeled with their mode number.
}
\label{cav}
\end{figure}

\begin{figure}
\caption{
Hyperfine measurements. In (a), we compare our value of the 
$5P_{1/2}$ hyperfine constant $A$ in $^{85}$Rb to earlier 
values from Arimondo et al.\ [5] and Barwood et al.\ [4]. The 
inset shows the hyperfine shifts inferred from our 
measurements. In (b), we show analogous results for $^{87}$Rb.
}
\label{hyper}
\end{figure}

\begin{table}
\caption{ 
The table lists the measured frequencies of the various 
transitions. The errors are statistical ($1 \sigma$) 
deviations. Measurements 1 and 2 are independent measurements 
of the same transition using different hyperfine transitions 
for the reference laser. They act as a check on systematic 
errors. Similarly, 3 and 4 are independent measurements. 
Measurements 7 and 8 are measurements of the $D_2$ line in 
$^{85}$Rb whose difference should be 31.705(50) MHz.
}
\begin{tabular}{lc}
\multicolumn{1}{c}{Measured transition} & Frequency (MHz) \\
\tableline
1. $D_1$ line: $^{85}{\rm Rb},F=3 \rightarrow F'=3 $ & 377 106 
271.606(13) \\
2. $D_1$ line: $^{85}{\rm Rb},F=3 \rightarrow F'=3 $ & 377 106 
271.560(21) \\
3. $D_1$ line: $^{85}{\rm Rb},F=3 \rightarrow F'=(2,3) $ & 377 
106 090.542(16) \\
4. $D_1$ line: $^{85}{\rm Rb},F=3 \rightarrow F'=(2,3) $ & 377 
106 090.601(22) \\
5. $D_1$ line: $^{87}{\rm Rb},F=2 \rightarrow F'=1 $ & 377 104 
392.053(21) \\
6. $D_1$ line: $^{87}{\rm Rb},F=2 \rightarrow F'=2 $ & 377 105 
205.093(18) \\
7. $D_2$ line: $^{85}{\rm Rb},F=3 \rightarrow F'=3 $ & 384 229 
121.099(14) \\
8. $D_2$ line: $^{85}{\rm Rb},F=3 \rightarrow F'=(2,3) $ & 384 
229 089.320(19) \\
\end{tabular}
\label{freqs}
\end{table}

\mediumtext
\begin{table}
\caption{ 
The table lists the frequencies of the $D_1$ line, $D_2$ line, 
and fine-structure interval in the two isotopes of Rb. Also 
given are the isotope shifts. The values are obtained from an 
analysis of the data in Table I and the errors are a 
combination of statistical and systematic errors.}
\begin{tabular}{cddd}
& \multicolumn{1}{c}{$^{85}$Rb (MHz)} & 
\multicolumn{1}{c}{$^{87}$Rb (MHz)} 
& Isotope shift (MHz) \\
\tableline
$D_1$ line & 377 107 385.623(50) & 377 107 463.209(50) & 
77.586(70) \\
$D_2$ line & 384 230 406.528(50) & 384 230 
484.468(10)\tablenote{From Ref 2}& 77.940(50) \\
F.S.\ Interval & 7 123 020.905(70) & 7 123 021.259(50) & 
0.354(90) \\
\end{tabular}
\label{values}
\end{table}

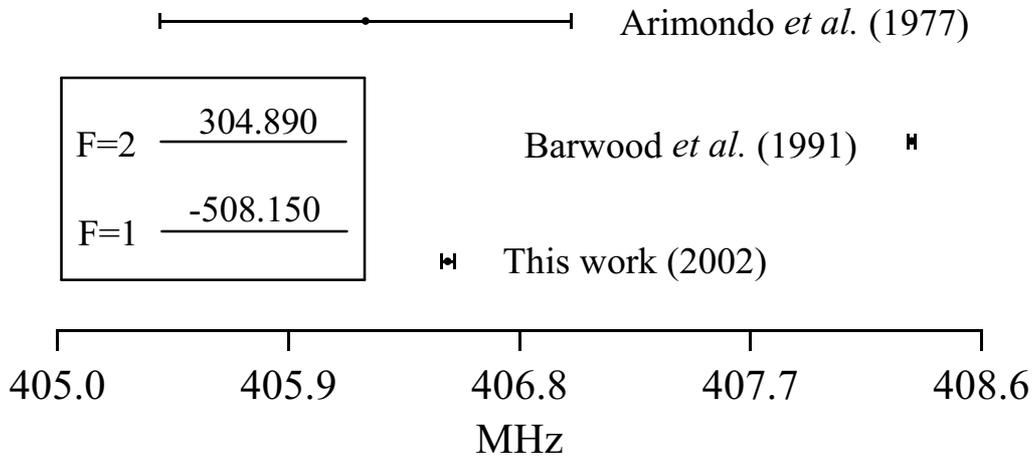
\end{document}